\documentclass[reprint, aps,prd,twocolumn,showpacs,showkeys, 10pt, longbibliography, nolinenumbers, floatfix]{revtex4-1}

\usepackage{amsmath}
\usepackage{graphicx}
\usepackage{float}
\usepackage{dcolumn}
\usepackage{multirow}
\usepackage{natbib}
\usepackage{bm}
\usepackage{changes}
\usepackage{comment}
\usepackage[colorlinks = true,linkcolor = blue,urlcolor  = blue,citecolor = blue,anchorcolor = blue]{hyperref}
\usepackage{enumerate}

\begin{document}

\title{Speed of sound bounds and first-order phase transitions in compact stars}

\author{P. Laskos-Patkos $^{1}$}
\email{plaskos@physics.auth.gr}

\author{G.A. Lalazissis $^{1}$}
\email{glalazis@auth.gr}

\author{Sibo Wang $^{2}$}
\email{sbwang@cqu.edu.cn}

\author{ Jie Meng $^{3}$}
\email{mengj@pku.edu.cn}

\author{Peter Ring $^{4}$}
\email{peter.ring@tum.de}

\author{Ch.C. Moustakidis $^{1}$}
\email{moustaki@auth.gr}

\affiliation{$^{1}$ Department of Theoretical Physics, Aristotle University of Thessaloniki, 54124 Thessaloniki, Greece}

\affiliation{$^{2}$ Department of Physics and Chongqing Key Laboratory for Strongly Coupled Physics, Chongqing University, Chongqing 401331, China}

\affiliation{$^{3}$ State Key Laboratory of Nuclear Physics and Technology, School of Physics, Peking University, Beijing 100871, China}

\affiliation{$^{4}$ Department of Physics, Technische Universit\"at M\"unchen, D-85747 Garching, Germany}

\begin{abstract}
In the present study, we employ three distinct, physically motivated speed of sound bounds to construct hybrid models, where the high-density phase is described by the maximally stiff equation of state. In particular, we consider the bounds related to special relativity, relativistic kinetic theory and conformality. The low-density hadronic phase is described by a state-of-the-art microscopic relativistic Brueckner-Hartree-Fock theory. This work aims to access the effect of the different speed of sound constraints on the relevant parameter space of the key parameters of first-order phase transitions by utilizing recent astronomical data. This involves a systematic analysis that also includes two distinct schemes for the construction of hybrid models (abrupt and smooth). Finally, a relevant discussion is conducted on the possible occurrence of a thermodynamic inconsistency that is related to the stability of the high-density phase over hadronic matter at large densities. 


\end{abstract}

\maketitle

\section{Introduction}

Neutron stars serve as unique natural laboratories for the physics of strongly interacting systems \cite{Bielich-2020}. Given that different models for the nuclear equation of state (EOS) predict distinct values for the bulk properties of neutron stars, then astronomical observations can be directly exploited for the imposition of constraints on the properties of ultradense matter \cite{Bielich-2020,Lattimer-2001}. Apart from static properties such as masses, radii, and tidal deformabilities, fruitful information may also be derived through the study of neutron star cooling~\cite{Page-2004,Page-2011,Lyra-2023}, oscillation modes~\cite{Flores-2014,Ranea-2018,Pradhan-2021,Pradhan-2024,Moustakidis-2015} and other dynamic phenomena.

A milestone in terms of our ability to estimate the main properties of neutron stars came with the NICER mission~\cite{Miller-2019,Riley-2019,Raaijmakers-2019,Vinciguerra-2024}. In particular, the analysis of soft x-ray pulses emitted from PSR J0030+0451 provided the first simultaneous mass and radius measurements for an isolated neutron star. According to the work of Riley {\it et al.}~\cite{Riley-2019}, the
mass and radius of PSR J0030+0451 are equal to $M=1.34^{+0.15}_{-0.16}M_\odot$ and $R=12.71^{+1.14}_{-1.19}$ km (in $1\sigma$). Furthermore, the relevant independent analysis of Miller {\it et al.}~\cite{Miller-2019} provided rather similar results. Interestingly, the recent study of Ref.~\cite{Vinciguerra-2024} revisited the analysis by incorporating additional data. By employing models similar to the aforementioned studies, the authors were able to replicate the initial estimation. However, they also derived a set of more complex solutions that may provide a better fit to the data. More precisely, the two obtained solutions suggest that either $M\sim1.4 M_\odot$, $R\sim11.5$ km or $M\sim1.7M_\odot$, $R\sim14.5$ km~\cite{Vinciguerra-2024}. Owing the very different estimations in the latter context, in this work we will rely on the initial estimation (specifically the one of Riley {\it et al.}~\cite{Riley-2019}, which provides a lower radius estimation). In any case, future studies will potentially shed light on this issue.

The recent analysis of Doroshenko {\it et al.}~\cite{Doroshenko-2022} on the mass and radius of the central compact object in the  HESS J1731-347 remnant suggests the interestingly small values of $M=0.77^{+0.20}_{-0.17}M_\odot$ and $R=10.4^{+0.86}_{-0.78}$km (in $1\sigma$). Such a measurement is of utmost importance as up to this moment the low-density part of the equation of state has been primarily constrained via nuclear experiments or theoretical calculations deriving from first principles and realistic many-body interactions~\cite{Akmal-1998,Gezerlis-2014,Lynn-2016,Tews-2018}. In addition, from an astrophysical perspective, it is rather unclear how a supernova explosion can lead to the formation of a such a light compact object~\cite{Suwa-2018}. As a consequence, a wave of studies has been triggered in the quest of unravelling the nature of this strangely light supernova remnant~\cite{DiClemente-2023,Horvath-2023,Oikonomou-2023,Das-2023,Rather-2023,Tsaloukidis-2023,Brodie-2023,Sagun-2023,Huang-2023,Li-2023,Kubis-2023,Routaray-2023,Laskos-2024,Li-2023n,Mariani-2024,Char-2024,Tewari-2024}. However, in spite of several arguments in favor of the aforementioned estimation~\cite{Kubis-2023}, it should be pointed out that there is still some debate on its validity~\cite{Alford-2023}.

Interestingly, the measurements on the HESS J1731-347 remnant~\cite{Doroshenko-2022} suggest that the nuclear equation of state should be rather soft at low densities, while the analyses of Riley {\it et al.}~\cite{Riley-2019} and Miller {\it et al.}~\cite{Miller-2019} on the mass and radius of PSR J0030+0451 indicate a stiffer behavior for nuclear matter at intermediate density (when considering the $1\sigma$ estimations). As pointed out by Kubis {\it et al.}~\cite{Kubis-2023} the simultaneous explanation of both measurements would require a 'Z-like' shape for the $M$-$R$ curve. In other words it is essential that the mass-radius dependence of compact stars is characterized by a region where $dM/dR>0$, which resembles the stiffening of the EOS. This is of particular importance as the stiffness of an EOS is directly connected to the speed of sound in dense matter ($c_s$). Therefore, we could potentially exploit the aforementioned measurements to gain some insight concerning one of the key ingredients that characterizes the nuclear EOS.

In the past decades, there has been extensive dialogue and research on the determination of a possible upper speed of sound bound~\cite{Hartle-1978,Lattimer-2014,Bedaque-2015,Olson-2000,Douchin-2001,Hippert-2024}. Considering the theory of special relativity, it is straightforward that $c_s$ should not exceed the speed of light ($c$). However, as pointed out by Lattimer~\cite{Lattimer-2014}, the causality constraint may be too extreme due to the fact that at asymptotically high-density $c_s$ should approach $c/\sqrt{3}$ from below (conformal limit). Furthermore, Bedaque and Steiner~\cite{Bedaque-2015} used simple arguments to support the validity of $c/\sqrt{3}$ as the upper speed of sound bound in the interior of neutron stars. Nonetheless, they demonstrated that current knowledge on the low-density sector of the nuclear EOS and the observation of massive compact stars disfavor the conformal constraint~\cite{Bedaque-2015}. Finally, it is worth mentioning that the condition $c_s<c$ is not necessarily  sufficient for causality in neutron star matter. As indicated by Olson~\cite{Olson-2000}, stringent constraints on causality can only be imposed in the framework of relativistic kinetic theory which describes all of the modes that can propagate in dense matter~\cite{Olson-2000,Douchin-2001}.

The main objective of the present work is the simultaneous reconciliation of HESS J1731-347 and PSR J0030$+$0451 in the framework of hybrid stars (with a maximally stiff high-density phase). As previously mentioned, the explanation of these two observational constraints requires a sufficiently stiff EOS. Hence, if one considers that there is an upper limit on the speed of sound, then there is also a bound on how stiff the EOS can be. Therefore, there is, first, a strong incentive to clarify how the imposition of different, physically motivated speed of sound limits may affect the ability of hybrid theoretical models to accurately reproduce the measurements mentioned above.

Secondly, the considered methodology of constructing hybrid EOSs with a maximally stiff high-density phase, sets the ground for the extraction of robust constraints on the main properties of first-order phase transitions (i.e., transition density $n_{\rm tr}$ and energy density jump $\Delta\mathcal{E}$). This is fairly simple to understand as the maximally stiff EOS is going to predict the least compact stellar configuration for a given stellar mass (compared to softer models with the same transition density and density jump). As a consequence, if for a given set of $n_{\rm tr}$ and $\Delta\mathcal{E}$ (which yield results compatible with HESS J1731-347) the maximally stiff EOS does not suffice to explain PSR J0030$+$0451, then there is no different high-density phase parametrization that could do so. 

Lastly, to enrich and extend our study, we do not only investigate sharp phase transitions (derived via Maxwell construction), which is the most frequent scenario found in the literature, but we also study the case where there is a crossover between the two phases (mimicking EOSs derived via Gibbs construction).~Our main motivation is to examine how the stiffness of the crossover region may affect the reconciliation of the considered astrophysical constraints. 

This paper is organized as follows. Section~\ref{secII} is devoted to the description of the hadronic model used. In Sec.~\ref{secIII} we present the methodology associated with the construction of hybrid EOSs with a maximally stiff description for the high-density phase. Section~\ref{secIV} contains a discussion of our results and their implications. Finally, Sec.~\ref{secV} contains a summary of our findings.

\section{Hadronic matter}~\label{secII}
We begin with an equation of state for hadronic matter, obtained from a state-of-the-art $ab$ $initio$ method, the self-consistent relativistic Brueckner-Hartree-Fock (RBHF) theory. It starts with a microscopic nucleon-nucleon interaction, the relativistic version of the potential Bonn A. Using the Thompson approximation~\cite{Thompson1970-PRD1-110} to the fully relativistic four-dimensional scattering equation, the Bethe-Salpeter equation, this bare nucleon-nucleon force has been determined by adjusting in Refs.~\cite{Machleidt1989_ANP19-189,Brockmann1990_PRC42-1965} the parameters of a one-boson-exchange potential (OBEP) with high precision to the experimentally determined scattering phase shifts. Since we are using a relativistic two-body interaction we do not need a microscopic three-body force, which, in the nonrelativistic case, is always a source of additional phenomenological parameters and additional uncertainties~\cite{Anastasio1980_PRL45-2096,Anastasio1981_PRC23-2273,Anastasio1983_PR100-327,Song1998_PRL81-1584}.

In the next step, the relativistic nuclear many-body problem of symmetric nuclear matter is treated using the Brueckner-Hartree-Fock approximation. This is a mean-field approximation based on an effective two-body interaction. The bare nucleon-nucleon force $V$ is much too strong at short distances to allow for a mean-field calculation. Therefore, an effective two-body interaction, the $G$-matrix, is used in the Brueckner-Hartree-Fock method. It takes into account that a scattering process in the medium does not allow a scattering to states below the Fermi surface, which are occupied in the medium.

To determine this effective interaction, the Thompson equation has to be solved with the Pauli operator $Q$, which excludes scattering to occupied states. In practice, the determination of this interaction needs a two-fold iteration process: Starting with initial single particle potentials, first the scattering equation in the medium, the Bethe-Goldstone equation has to be solved:
\begin{equation}\label{eq:ThomEqu}
G(W)=\ V~+V\frac{Q}{W-H_0 + i\epsilon}G(W),
\end{equation}
where $W$ is the starting energy and $H_0$ describes the independent motion of the two particles in the mean-field.

In the final step, this $G$-matrix has to be used for a relativistic Hartree-Fock calculation~\cite{Brockmann1978_PRC18-1510} and the determination of the updated single-particle potentials. This iteration has to be continued until convergence is achieved.

Solving this double iteration process~\cite{WANG-SB2021_PRC103-054319, Wang-2022a} for each density $n$, one can calculate the energy per particle $\cal E$ and finds the equation of state ${\cal E}(n)$ used in the following sections. For a broader analysis on the application of the RBHF theory for the study of compact star properties the reader is referred to Refs.~\cite{Tong-2022,Wang-2022b,Qu-2023,Qin-2023,Farrell-2024}.

\section{Hybrid equation of state}~\label{secIII}

As previously mentioned, the main objective of the present study relies on the construction
of hybrid EOSs, where the high-density phase is characterized  as maximally stiff. Considering that the stiffness of an EOS is directly related to the speed of sound, in this work we consider three distinct, physically motivated bounds. In particular, we consider the limits: 
\begin{enumerate}
    \item $\frac{c_s}{c}\leq1$, from special relativity (causality limit)~\cite{Hartle-1978}.
    \item $\frac{c_s}{c}\leq \frac{1}{\sqrt{3}}$, from QCD and other theories~\cite{Bedaque-2015}.
    \item $\frac{c_s}{c} \leq \left( \frac{\mathcal{E}-P/3}{\mathcal{E}+P}\right)^{\frac{1}{2}}$, from relativistic kinetic theory~\cite{Olson-2000}.
\end{enumerate}
It is worth commenting that all of the aforementioned bounds have been employed numerous times in the literature in an attempt of studying the properties of maximally massive neutron stars~\cite{Bedaque-2015,Olson-2000,Moustakidis-2016,Margaritis-2020,Kanakis-2020,Reed-2020,Nath-2002,Roy-2024}.

Apart from the assumption on a maximal speed of sound value, the construction of a hybrid EOS requires additional physical input.~In particular, depending on the number of conserved charges between the two phases, we can follow two distinct methods~\cite{Bielich-2020,Glendenning-1992,Costantinou-prd}. In the case of a single conserved charge (baryon number) the associated approach is known as Maxwell construction. In this case, the phase transition is abrupt resulting in a discontinuity in the energy density [and we will refer to it as abrupt construction (AC)]. In the scenario of multiple conserved charges (baryon number and electric charge) the appropriate treatment is the so-called Gibbs construction. Contrary to the Maxwell construction where the two phases coexist only for a single value of pressure, under the Gibbs framework the EOS is continuous 
and the two phases form a mixed phase in the stellar interior. Since in the present study we will be working with a phenomenological model that only mimics the existence of a mixed phase (i.e., we will use a polytropic parametrization for the description of the crossover) we are going to refer to the continuous construction as smooth construction (SC).

Let us now present the ansantz describing the derivation of hybrid EOSs. It is important to comment that the description related to the first two bounds is going to be different than the one concerning relativistic kinetic theory, since in the latter case the maximum speed of sound value is density dependant. We begin by presenting the framework related to the constant speed of sound bounds (i.e., $c_s^2\leq c^2$ and $c_s^2\leq c^2/3$). In the case of AC the resulting hybrid EOS reads as~\cite{Alford-2013,Montana-2019} 

\begin{equation}
  \mathcal{E}(P) = \begin{cases} 
      \mathcal{E}_{\rm HADRON}(P), & P\leq P_{\rm tr} \\
      \mathcal{E}(P_{{\rm tr}}) + \Delta \mathcal{E} + (c_s/c)^{-2}(P-P_{{\rm tr}}), & P > P_{{\rm tr}},
   \end{cases}
   \label{1}
\end{equation}
where $P_{\rm tr}$ and $ \Delta \mathcal{E}$ denote the phase transition pressure and the energy density discontinuity (or jump), respectively. It is important to comment that the first line of Eq.~(\ref{1}) refers to the hadronic phase (presented in Sec.~\ref{secII}), while the second one to the maximally stiff high-density phase. In the case of SC we have~\cite{Montana-2019}

\begin{equation}
  \mathcal{E}(P) = \begin{cases} 
      \mathcal{E}_{\rm HADRON}(P), & P\leq P_{\rm tr} \\
      A_m (P/K_m)^{1/\Gamma_m}+ P / (\Gamma_m-1), & P_{\rm tr} < P\leq P_{{\rm CSS}} \\
      \mathcal{E}(P_{\rm CSS}) + (c_s/c)^{-2}(P-P_{{\rm CSS}}), & P > P_{{\rm CSS}},
   \end{cases}
   \label{2}
\end{equation}
where now the second line denotes the presence of a crossover region (mimicking the mixed phase) which is modeled via a polytopic EOS of the form $P(n)=K_m n^{\Gamma_m}$. Note that the parameters $A_m$ and $K_m$ are evaluated by imposing continuity of the EOS at $P_{\rm tr}$. Finally, $P_{{\rm CSS}}$ stands for the pressure at which the mixed phase ends and the constant speed of sound (CSS) region starts. 

One of the objectives of the present work is to study how the stiffness of the crossover region may affect the reconciliation of HESS J1731-347 and PSR J0030$+$0451. Notably, in the current model, the stiffness of this density regime is regulated via the polytropic index $\Gamma_m$. Hence, we aim to examine how the viable (i.e, compatible to observations) $n_{\rm tr}$, $\Delta\mathcal{E}$ sets change when varying $\Gamma_m$. Owing the fact that the explanation of HESS J1731-347 may require a sufficiently strong phase transition (soft enough crossover), we are going to employ a starting $\Gamma_m$ value of 1.03, which has been employed in previous works for the construction of twin stellar configurations~\cite{Tsaloukidis-2023,Montana-2019}.

For comparison reasons we wish anchor our EOS description on the same parameters, regardless of whether we employ AC or SC. For that matter, the two free parameters in both cases are the energy density jump and the transition pressure (or density). However, no energy density discontinuity appears in SC, and hence we identify the value of $\Delta\mathcal{E}$ as the range of the crossover region (which enables the evaluation of $P_{{\rm CSS}}$)~\cite{Montana-2019}.

Let us now proceed to the description of hybrid EOSs in the case where we consider the maximal speed of sound value as it was derived by Olson in the framework of relativistic kinetic theory~\cite{Olson-2000}. It should be stressed out that this limit is only approximate since its exact derivation requires precise knowledge for the viscosity of dense matter. However, it is still particularly useful to employ it as a less extreme causality limit compared to the one provided by special relativity. Following Refs.~\cite{Olson-2000,Moustakidis-2016,Margaritis-2020},  the maximally stiff EOS will be derived through the solution of the ordinary differential equation
\begin{equation} \label{3}
    \frac{dP}{d\mathcal{E}}= \frac{\mathcal{E}-P/3}{\mathcal{E}+P}.
\end{equation}
By considering standard thermodynamic relations, Eq.~(\ref{3}) can be rewritten as
\begin{equation}
    \label{4}
    n^2\frac{d^2\mathcal{E}}{dn^2}+\frac{n}{3}\frac{d\mathcal{E}}{dn}-\frac{4}{3}\mathcal{E}=0,
\end{equation}
which can be easily solved analytically and leads to
\begin{equation} \label{5}
    \begin{split}
        \mathcal{E}(n) & =\mathcal{C}_1 n^{a_1}+\mathcal{C}_2 n^{a_2}, \hspace{3 mm} a_1=\frac{1+\sqrt{13}}{3}, \\
        &a_2=\frac{1-\sqrt{13}}{3},
    \end{split}
\end{equation}
and
\begin{equation} \label{6}
    P(n) =\mathcal{C}_1 n^{a_1} (a_1-1)+\mathcal{C}_2 n^{a_2}(a_2-1).
\end{equation}
Therefore, the resulting hybrid EOSs will be derived as presented through Eqs.~(\ref{1}) and (\ref{2}), with the modification that their last lines will now be replaced with the EOS described through Eqs.~(\ref{5}) and~(\ref{6}). The constants of integration $\mathcal{C}_1$ and $\mathcal{C}_2$ can be easily derived by considering the values of $P$ and $\mathcal{E}$ at the transition point which are equal to $P_{\rm tr}$ and $\mathcal{E}_{\rm tr}+\Delta\mathcal{E}$, respectively.

\section{Results and Discussion}~\label{secIV}

\subsection{Effects on the maximum mass}~\label{secIVa}

\begin{figure*}[t]
  \centering  \includegraphics[width=\linewidth,scale=1]{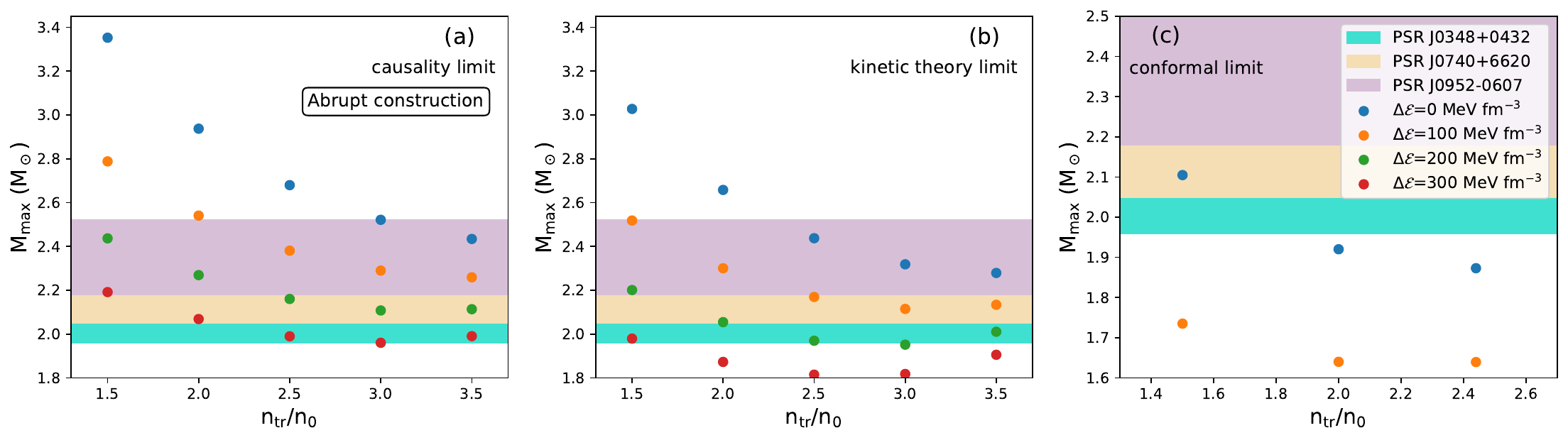}
  \caption{The maximum mass as a function of the transition density for different speed of sound bounds. Hybrid EOSs were constructed via AC. In panels (a),(c) the upper bounds on the sound velocity are $c$ and $c/\sqrt{3}$, respectively. In panel (b) the speed of sound is constrained via relativistic kinetic theory. The shaded regions correspond to possible constraints on the maximum mass from the observation of PSR J0348$+$0432~\cite{Antoniadis-2013}, PSR J0740$+$6620~\cite{Cromatie-2020}, and PSR J0952-0607~\cite{Romani-2022}.}
  \label{Mmax_m}
\end{figure*}
\begin{figure*}[t]
  \centering  \includegraphics[width=\linewidth,scale=1]{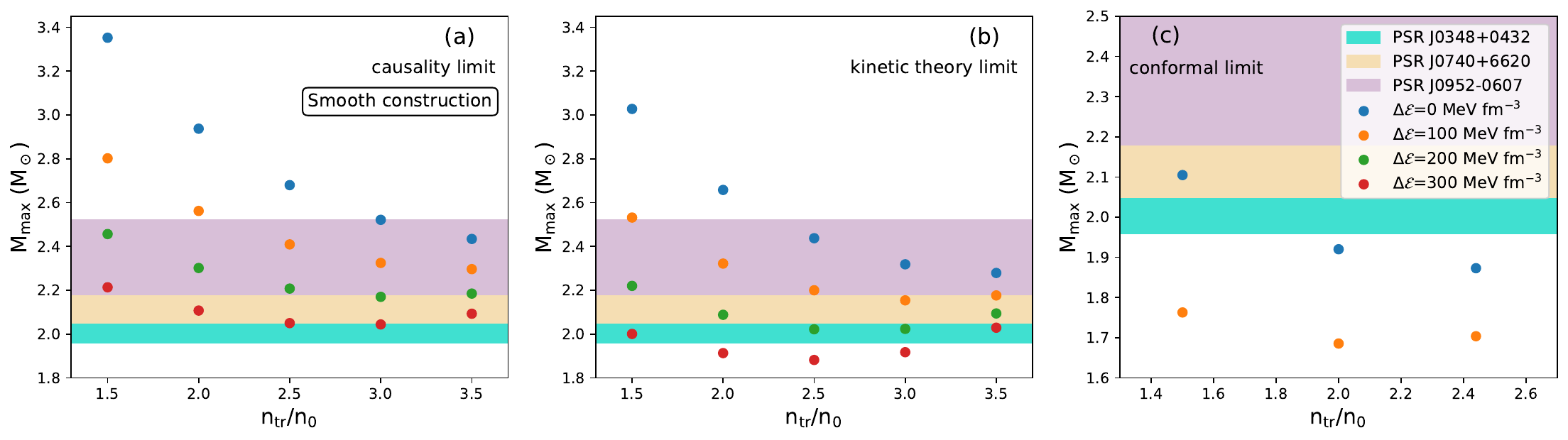}
  \caption{Same as Fig.~\ref{Mmax_m} but for the case were the hybrid EOSs were constructed via SC. Note that, $\Delta\mathcal{E}$ (appearing in the legend) does not refer to a discontinuity but to the width of the crossover region. The polytropic index is equal to 1.03.}
  \label{Mmax_g}
\end{figure*}

Some of the most robust constraints regarding our knowledge on neutron star structure come from the observation of massive compact stars.  Note that for a given set of $n_{\rm tr}$ and $\Delta\mathcal{E}$ values the maximally stiff EOS predicts the largest possible maximum mass. Therefore, we begin our analysis by addressing the effects of different speed of sound bounds on the maximum mass of hybrid compact stars. More precisely, we have constructed sets of hybrid EOSs by varying the transition density and the energy density discontinuity in the ranges $[1.5n_0,3.5n_0]$ and $[0,300]$ MeV fm$^{-3}$, respectively ($n_0$ denotes the nuclear saturation density). It is important to comment that in the case of $c_s\leq c/\sqrt{3}$ the maximum considered  transition density is $\sim2.5n_0$, since beyond this value the speed of sound in the hadronic phase exceeds the conformal limit.

Figures~\ref{Mmax_m} and~\ref{Mmax_g}
depict the maximum mass as a function of the transition density for AC and SC, respectively. The shaded regions correspond to possible constraints on the maximum mass from the observation of PSR J0348+0432~\cite{Antoniadis-2013}, PSR J0740+6620~\cite{Cromatie-2020} and PSR J0952-0607~\cite{Romani-2022}. Furthermore,
the effect of different energy density jump values is also included. First of all, it is clear that the selection of the construction method (AC or SC) does not appear to play a critical role concerning the resulting maximum mass prediction. In particular, the use of a smooth crossover (with $\Gamma_m=1.03$) results into an increase of less than$\sim0.1 M_\odot$. However, the resulting maximum mass is tightly connected to the selection of the polytropic index that controls the EOS of the crossover region. A larger $\Gamma_m$ value would in principle result into larger deviations between the AC and SC predictions,  as the EOS constructed with the latter method would become stiffer. 

Another remark that can be made concerns the trend of the $M_{\rm max}(n_{\rm tr})$ curve.~In particular, the case of $\Delta\mathcal{E}=0$~MeV fm$^{-3}$ has been considered numerous times in the literature and it indicates that the maximum mass decreases as a function of the transition density~\cite{Moustakidis-2016}. However, Figs.~\ref{Mmax_m} and~\ref{Mmax_g} indicate that the inclusion of a softening region (density jump for AC and crossover for SC) in the EOS can invert this behavior. More precisely, for a given (non zero) $\Delta\mathcal{E}$ the value of $M_{\rm max}$ initially decreases with $n_{\rm tr}$ and after a certain density value it starts to increase. This could be explained by the fact that the case where the hadronic model is matched directly and continuously with the maximally stiff EOS is fundamentally different from the case where an energy density jump (or crossover) is present.~In the first scenario, the earlier the transition, the earlier the EOS becomes maximally stiff. In contrast, the inclusion of an energy density discontinuity (or crossover) initially softens the EOS and this has an important effect on the resulting maximum mass. For instance,
the softening introduced in the case of an early phase transition can result into a maximum mass prediction that is lower compared to the mass of the most massive purely hadronic configuration when a larger transition density is considered.

Finally, it is important to comment that the inclusion of a softening region in the case where the maximal speed of sound is set to $c/\sqrt{3}$ leads to predictions that are incompatible with the observation of massive compact stars that exceed $2M_\odot$. More massive neutron stars like PSR J0952-0607 ($\sim 2.35 M_\odot$) are in strong tension with the conformal bound, regardless of the selected energy density jump. This enhances the conclusion of previous works, which by employing generic, model independent EOSs have indicated that the sound velocity most likely exhibits a peak in the neutron star interior that surpasses the value $c/\sqrt{3}$~\cite{Tews-2018,Altiparmak-2022,Ecker-2022,Brandes-2024}.

\subsection{Effects on the $n_{\rm tr}-\Delta\mathcal{E}$ parameter space}

\begin{figure*}[t]
  \centering  \includegraphics[width=\linewidth,scale=1]{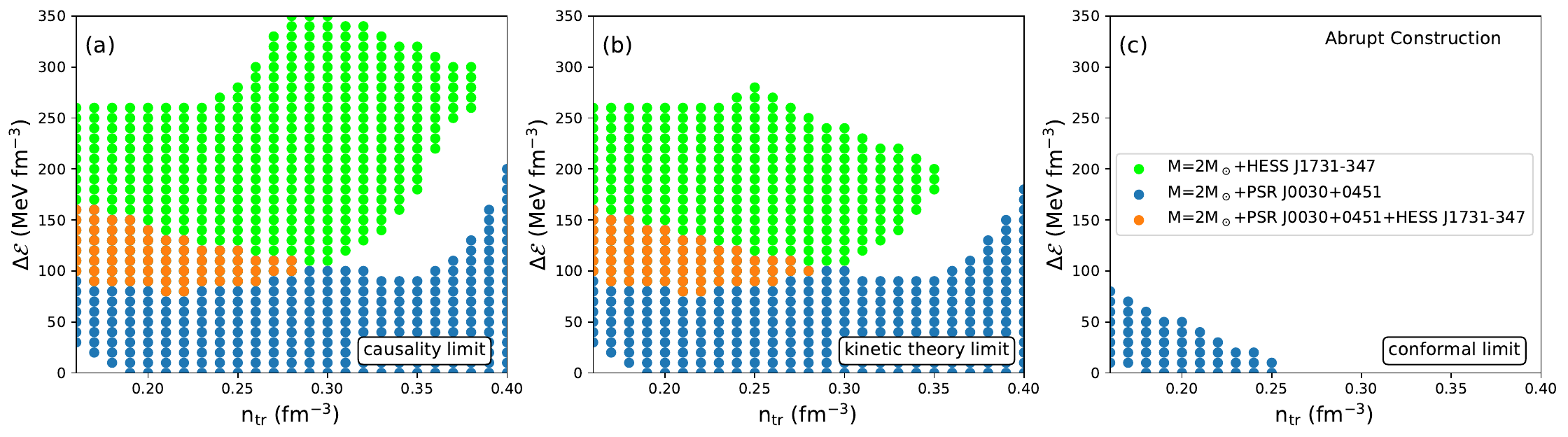}
  \caption{The $n_{\rm tr}-\Delta\mathcal{E}$ parameter space derived for the explanation of different astronomical measurements. Hybrid EOSs were constructed via AC. In panels (a),(c) the upper bounds on the sound velocity are $c$ and $c/\sqrt{3}$, respectively. In panel (b) the speed of sound is constrained via relativistic kinetic theory. The green shaded area corresponds to the valid parametrizations when the HESS J1731-347 constraints are considered, while the blue region indicates the parametrizations that are compatible with the mass and radius of PSR J0030+0451. The orange area illustrates the region where both of the aforementioned measurements are reproduced simultaneously. All of the depicted parametrizations yield EOSs that are compatible with the existence of massive compact stars that exceed $2M_\odot$. }
  \label{params-m}
\end{figure*}

\begin{figure*}[h]
  \centering  \includegraphics[width=\linewidth,scale=1]{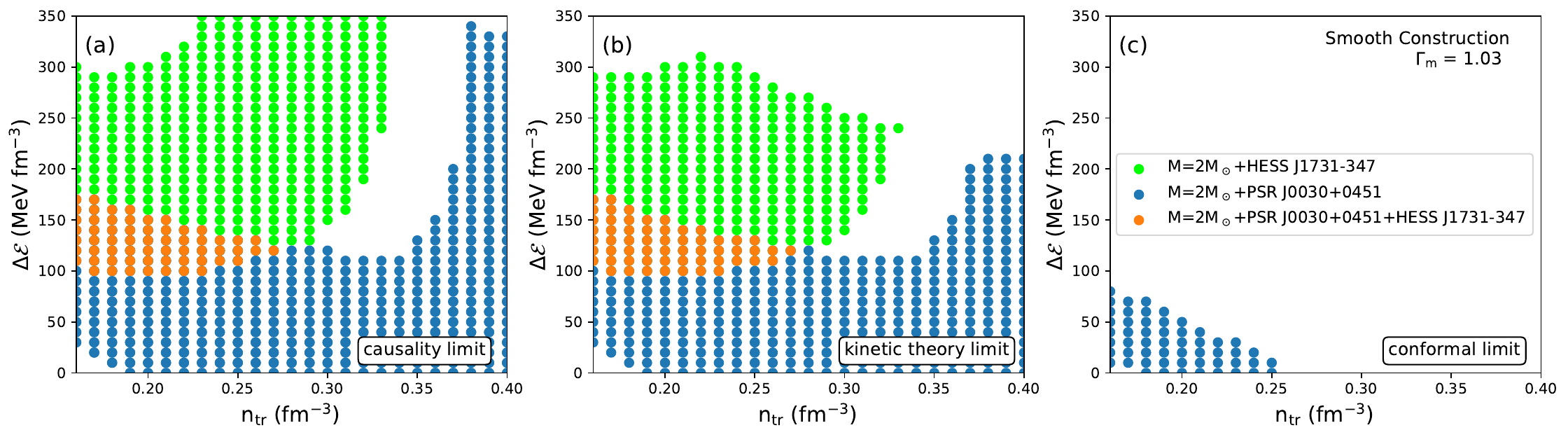}
  \caption{Same as Fig. \ref{params-m} but for the case were the hybrid EOSs were constructed via SC. Note that in the case of SC, $\Delta\mathcal{E}$ does not refer to a discontinuity but to the width of the crossover region. The polytropic index is equal to $\Gamma_m = 1.03$.}
  \label{params-g}
\end{figure*}

\begin{figure*}[t]
  \centering  \includegraphics[width=\linewidth,scale=1]{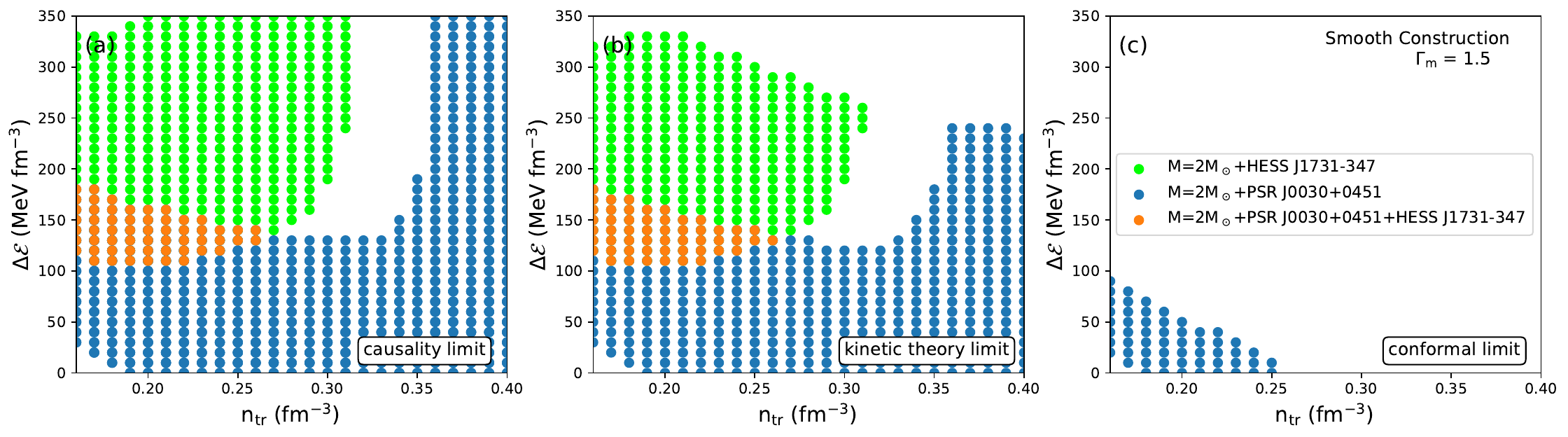}
  \caption{Same as Fig. \ref{params-m} but for the case were the hybrid EOSs were constructed via SC. Note that in the case of SC, $\Delta\mathcal{E}$ does not refer to a discontinuity but to the width of the crossover region. The polytropic index is equal to $\Gamma_m = 1.5$.}
  \label{params-g-2}
\end{figure*}

\begin{figure*}[t]
  \centering  \includegraphics[width=\linewidth,scale=1]{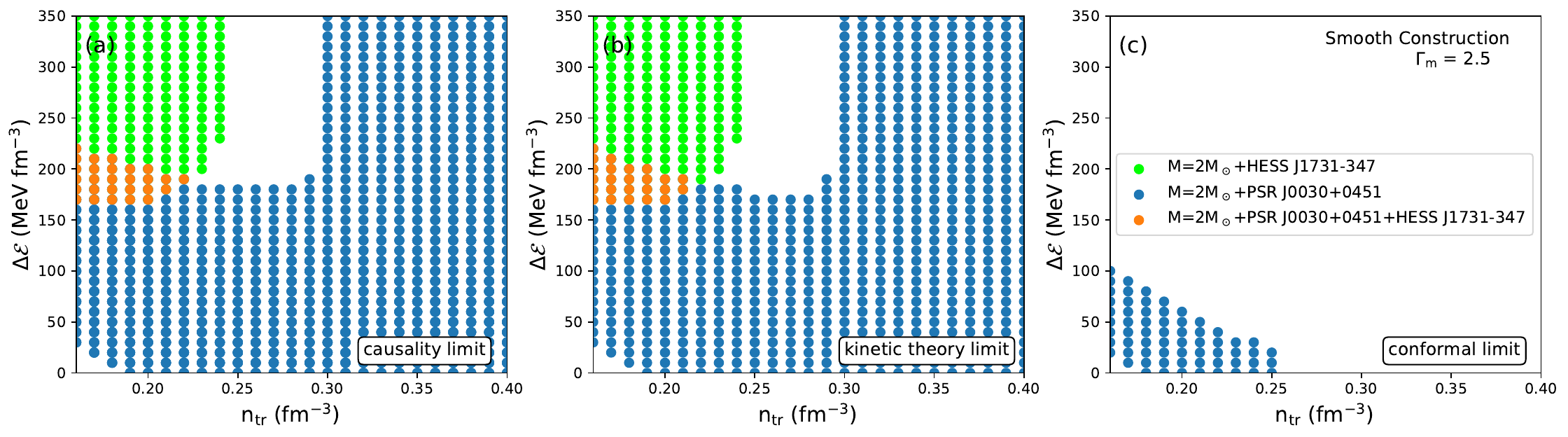}
  \caption{Same as Fig. \ref{params-m} but for the case were the hybrid EOSs were constructed via SC. Note that in the case of SC, $\Delta\mathcal{E}$ does not refer to a discontinuity but to the width of the crossover region. The polytropic index is equal to $\Gamma_m = 2.5$.}
  \label{params-g-3}
\end{figure*}

\begin{figure*}[t]
  \centering  \includegraphics[width=\linewidth,scale=0.5]{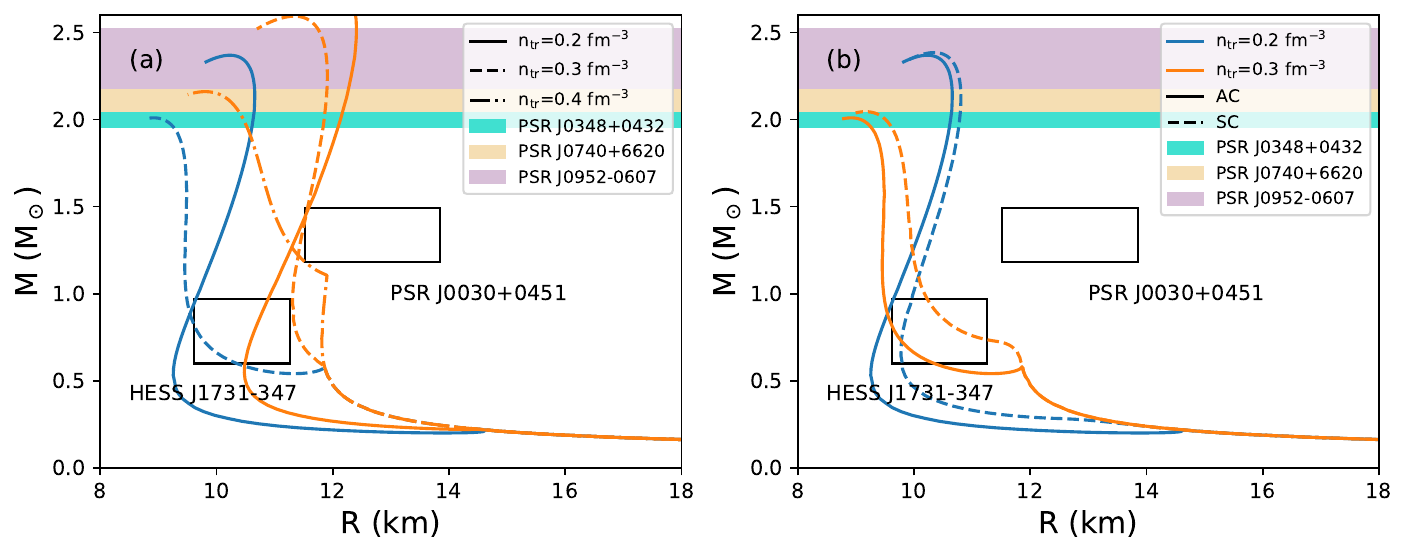}
  \caption{Panel (a): Mass-radius dependence for hybrid EOSs with transition densities $n_{\rm tr}=0.2,0.3,0.4$ fm$^{-3}$. The energy density jump takes the maximum possible value that leads to agreement with the HESS J1731-347 (blue curves) and PSR J0030+0451 (orange curves). Panel (b): Comparison of the mass-radius dependence for EOSs derived with AC (solid curves) and SC for $\Gamma_m=1.03$ (dashed curves). For the blue (orange) curves the transition density is $n_{\rm tr}=0.2$ ($0.3$) fm$^{-3}$ and the energy density jump is $\Delta\mathcal{E}=260$ ($350$) MeV fm$^{-3}$. Note that the term energy density jump in the case of SC does not refer to a discontinuity but to the width of the crossover region}.
  \label{MR}
\end{figure*}

\begin{figure}[h]
  \centering  \includegraphics[width=\linewidth,scale=1]{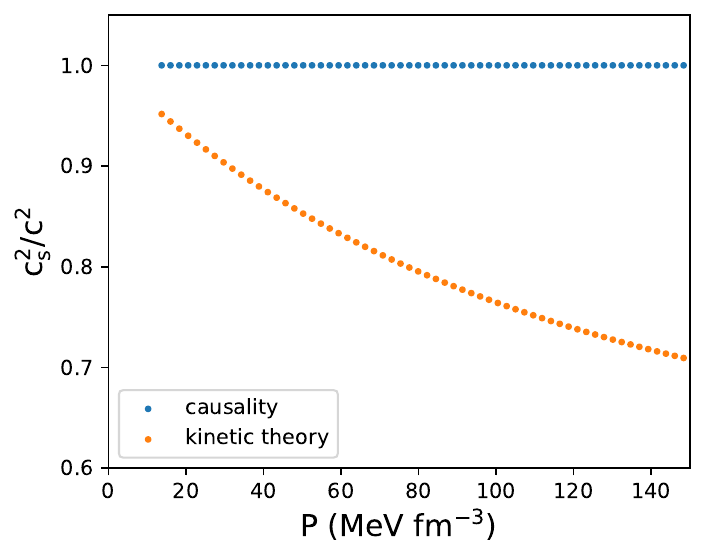}
  \caption{Sound speed for hybrid EOSs (constructed via AC) with $n_{\rm tr}=0.25$ fm$^{-3}$ and $\Delta\mathcal{E}=100$ MeV fm$^{-3}$ constrained by special relativity (blue) and relativistic kinetic theory (orange). The curves indicate the speed of sound at the high-density phase (i.e., after the phase transition).}
  \label{soundspeed}
\end{figure}

In this section, we aim to address which values of the free parameters $n_{\rm tr}$ and $\Delta\mathcal{E}$ allow for a simultaneous explanation of the masses and radii for HESS J1731-347 and PSR J0030$+$0451 in $1\sigma$. For each respective speed of sound bound we have varied the values of the transition density and the energy density jump in the ranges $[0.16,0.4]$ fm$^{-3}$ and $[0,350]$ MeV fm$^{-3}$. Then, the resulting sets of EOSs were applied to the Tolman-Oppenheimer-Volkov equations which provided the corresponding stellar models.

Figures~\ref{params-m}-\ref{params-g-3} illustrate the viable parametrizations when different astronomical measurements are considered. More precisely, Fig.~\ref{params-m} contains the results derived using AC, while Figs.~\ref{params-g},~\ref{params-g-2} and~\ref{params-g-3} correspond to the SC case with $\Gamma_m=1.03$, $1.5$ and $2.5$, respectively. The blue dots indicate the parameter values for which the derived EOSs satisfy the PSR J0030$+$0451 constraints ($1\sigma$) and also predict stable stellar configurations that exceed $2M_\odot$.~Let us begin our analysis for the bounds related to special relativity and relativistic kinetic theory.~It is worth noting that for EOSs constructed via AC, the highest possible values for the energy density jump are determined by the lowest estimation for the radius of PSR J0030$+$0451. The latter can be seen for a variety of different transition densities in Fig.~\ref{MR}(a). More precisely, in deriving Fig.~\ref{MR}(a), the EOSs are selected so that, for a given transition density, the energy density jump is maximized while ensuring the results remain consistent with individual observations (observations are considered separately and not combined at this point). In the case of SC, for low transition density values, the maximal possible $\Delta\mathcal{E}$ is dictated by the radius of PSR J0030$+$0451 (as in the case of AC). However, the factor that determines the highest energy density jump for larger transition densities is the $2M_\odot$ demand. The exact transition density at which the $2M_\odot$ constraint becomes the relevant one (concerning the maximal $\Delta\mathcal{E}$ estimation) depends on the value of the polytropic index. In particular, the larger the value of $\Gamma_m$ the lower the transition density. Finally, when the conformal limit is imposed, the largest $\Delta\mathcal{E}$ is also determined by the $2M_\odot$ constraint in both AC and SC.

The green dots, Figs.~\ref{params-m}-\ref{params-g-3}, depict the parametrizations that are compatible with the constraints on the mass and radius of the ultralight compact object in the HESS J1731-347 remnant ($1\sigma$) and also allow the existence of stable massive stars ($M_{\rm max}>2M_\odot$). Once again, for low transition density, the maximum $\Delta\mathcal{E}$ value is derived by considering the lowest estimation on the radius provided by the HESS J1371-347 constraints.~However, for higher $n_{\rm tr}$ the definitive factor, concerning the highest possible $\Delta\mathcal{E}$ is the $2 M_\odot$ constraint [see  Fig.~\ref{MR}(a)].~Note that beyond a transition density value, the explanation of the HESS J1731-347 constraints is impossible regardless of the selected energy density jump. That is due to the fact that the softening needed would be so large that it would lead to the formation of stellar configurations which would be either ultimately unstable or they would never surpass the $2 M_\odot$.

As is evident, the use of different EOS construction methods (AC and SC) appears to have an effect on the resulting parameter spaces that correspond to the explanation of PSR J0030+0451 and HESS J1731-347 (blue and green areas, respectively). This influence is more clearly reflected as the polytropic index, controlling the stiffness of the crossover region, increases.~Notably, for the same transition density values, the SC allows for larger energy density jumps compared to the case of AC. In addition, the larger the value of $\Gamma_m$ the larger the corresponding estimation on the highest possible energy density jump.~This could be explained by the fact that for the same values of $n_{\rm tr}$ and $\Delta\mathcal{E}$, an EOS becomes stiffer as the polytropic index increases.~As a consequence, the resulting compact stars will be characterized by larger radii values (when compared to equal mass configurations derived through AC or SC with lower $\Gamma_m$).~In addition, the stiffer an EOS the larger its resulting maximum mass prediction [see Fig.~\ref{MR}(b) comparing the $M$-$R$ dependence for EOSs constructed via AC and GC with the same $n_{\rm tr}$ and $\Delta\mathcal{E}$)]. Owing our previous analysis on the factors that determine the highest possible energy density jump, it becomes clear that EOSs characterized by higher $\Gamma_m$ will fit the observations for larger $\Delta\mathcal{E}$ values.~ A final remark concerns the fact that when AC is used, the explanation of HESS J1731-347 is possible for larger transition density values.~This may be attributed to the abrupt nature of AC, which leads to sharper phase transitions.

The orange dots in Figs.~\ref{params-m}-\ref{params-g-3} indicate the parametrizations that allow for the explanation of both HESS J1731-347 and PSR J0030$+$0451 $1\sigma$ constraints (and also align with the existence of massive compact stars with $2M_\odot$). A first important comment is that in the case where the speed of sound bound is set to be $c/\sqrt{3}$ there is not a single parameter combination that yields compatible results with the aforementioned measurements.~The latter can be explained by the fact that a nonzero energy density jump is essential for the explanation of HESS J1731-347. However, as we have shown in Sec.~\ref{secIVa}, an EOS bound by the conformal limit and characterized by a non zero $\Delta\mathcal{E}$ is disfavored by the existence of $2M_\odot$ pulsars.~For the two remaining speed of sound bounds we observed that the resulting parameter spaces appear to be similar. In both cases, for AC, the transition density is constrained below $\sim 0.28$ fm$^{-3}$, while the energy density jump varies in the range $\sim80-170$ MeV fm$^{-3}$ (see Figs.~\ref{params-m}(a) and~\ref{params-m}(b)). The aforementioned similarity is related to the fact that for low values of pressure the bound set by relativistic kinetic theory is close to the speed of light. The latter is illustrated in Fig.~\ref{soundspeed}. As one can observe, the value of the speed of sound is quite close to the speed of light around the transition pressure and then reduces rapidly. This speed of sound drop manifests in the resulting maximum mass prediction (see Figs.~\ref{Mmax_m} and~\ref{Mmax_g}). 

With regards to effect of the EOS construction method, we found that the polytropic index may have important impact on the simultaneous reconciliation of HESS J1731-347 and PSR J0030$+$0451 measurements. In particular, as the crossover region becomes stiffer (i.e., as the polytropic index increases), the largest possible transition density, that is compatible to observations, decreases (see Figs.~\ref{params-m},~\ref{params-g},~\ref{params-g-2} and~\ref{params-g-3}). This is of utmost importance, as a phase transition close to nuclear saturation is not possible due to the fact that nuclei are known to be stable at that density regime. Hence, we could potentially obtain information on the maximal possible stiffness of the crossover region.~Another remark is that as $\Gamma_m$ increases, the viable (i.e., compatible to observations) energy density jump regime is shifted towards larger values.

Considering the inability of deriving the properties of dense matter via quantum chromodynamics (due the so-called "lattice sign problem"), the main drawback of studies working on compact star structure is that their results are at some level model dependant. As the main objective of the present study is to impose constraints on the properties of first order phase transitions, it is of utmost importance to discuss how our results (in particular, the derived orange shaded area in Figs.~\ref{params-m},~\ref{params-g},~\ref{params-g-2} and~\ref{params-g-3}) would be modified under the consideration of different models. 

Firstly, the selection of a maximally stiff EOS for the description of the high-density phase provides a rather general picture for the resulting parameter space, at least in terms of setting an upper bound on the energy density jump for a given transition density. This is quite simple to understand since the maximally stiff description of the high-density phase produces stellar configurations with the largest possible radius for a given mass (for a specific selection of $n_{\rm tr}$ and $\Delta\mathcal{E}$)~\cite{Sun-2023}. As previously mentioned, the largest value of the energy density jump for a given transition density is driven by the lowest estimation on the radius of PSR J0030+0451. Note that any softer description of the high-density phase would lead to lower radii predictions. Therefore, there is no way to model the high-density phase with a larger energy density jump (compared to the maximum one derived via the use of the maximally stiff parametrization) that could produce results compatible with the PSR J0030+0451. However, the lowest estimation for the energy density jump is not as general since it is derived by the highest radius estimation for the compact object in HESS J1731-347. Therefore, a softer description of the high-density phase (in terms of sound velocity) could produce results compatible with the analysis for HESS J1731-347 for lower $\Delta\mathcal{E}$ values.~Finally, the parameter space is obviously dependent on the stiffness of the low-density hadronic phase. For instance, the selection of a stiffer hadronic EOS would potentially lead to larger energy density jump values as the reconciliation of HESS J1731-347 constraints would require a stronger phase transition (since the hadronic branch in the mass-radius diagram would be shifted to larger radii). In contrast, the selection of a softer realisation for hadronic matter could lead to the explanation of the HESS J1731-347 constraints without the need for a large density jump as the hadronic branch in the mass-radius diagram would be characterised by lower radii. It is due to all of these uncertainties that we have chosen to anchor our study on a microscopic relativistic model that is derived using a realistic potential. In addition, the fact that transition density is constrained below $\sim 0.28$ fm$^{-3}$ increases the reliability of our results since we avoid extrapolating away from nuclear saturation density. However, it would be very interesting to revisit this study and examine (quantitatively) the effects of varying the stiffness of the low-density phase on the resulting $n_{\rm tr}-\Delta\mathcal{E}$ parameter space.

\subsection{Stability of the high-density phase}

\begin{figure}[t]
  \centering  \includegraphics[width=\linewidth,scale=1]{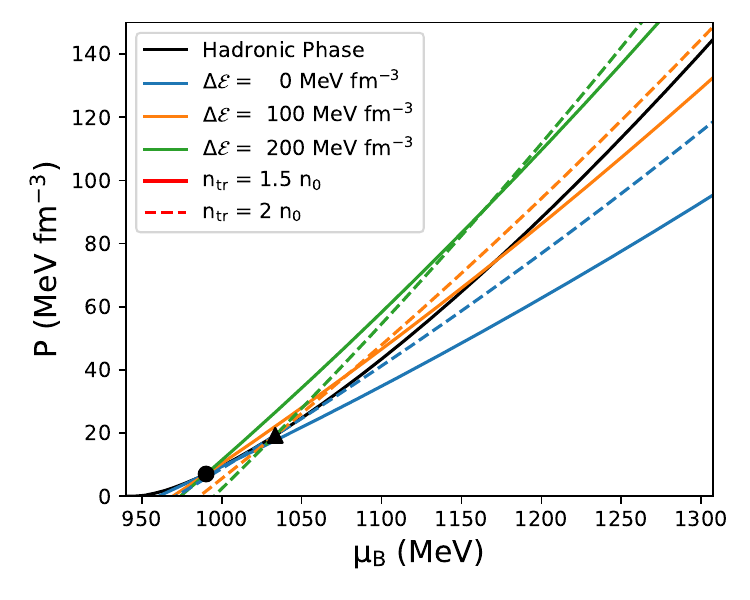}
  \caption{The pressure-baryon chemical potential relation for the hadronic EOS (black) and different constant speed of sound models ($c_s=c$). The solid (dashed) curves correspond to the case where $n_{\rm tr}=1.5n_0$ ($2n_0$). The circular (triangular) dot denotes the pressure and baryon chemical potential for $n=1.5n_0$ ($2n_0$).}
  \label{instability_pm}
\end{figure}

\begin{figure}[t]
  \centering  \includegraphics[width=\linewidth,scale=1]{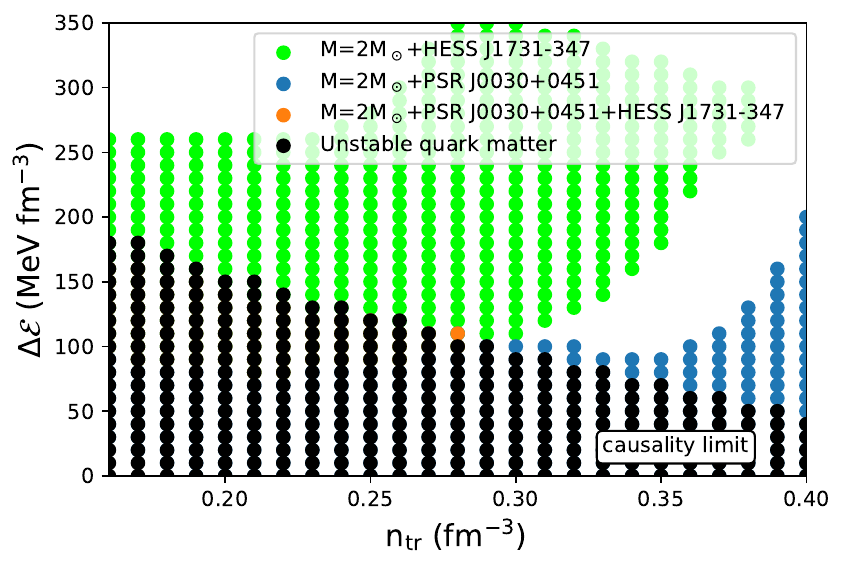}
  \caption{The parameter space presented in Fig.~\ref{params-m}(a). The black
circles denote the parametrizations for which} the hadronic phase becomes energetically favored again before $\mu_B=1307.25$ MeV.
  \label{strongPT}
\end{figure}

Hybrid EOSs (when studying hadron to quark phase transition) are often derived by the thermodynamically consistent combination of a well-constrained hadronic model with a quark
EOS. However, it is suggested that, based on several models, the speed of sound in quark matter (for densities relevant
to compact stars) can be approximated as a constant~\cite{Alford-2013}. The latter has led to a wide range of studies based on
the constant speed of sound parametrization for the modeling of quark matter~\cite{Alford-2014b,Christian-2019,Christian-2021,Christian-2022,Han-2019a,Li-2021,Sharifi-2021,Paschalidis-2018,Niseri-2024,Alford-2017,Deloudis-2021,Han-2020,Christian-2024,Li-2021,Li-2023h}. The main advantage of such
an approach is that the free parameters of the system are the transition density and the energy density
discontinuity. However, contrary to studies based on mean-field quark matter realizations, the stability of quark
matter (over hadronic matter) is not tested. This issue was first pointed out by Zdunik and Haensel~\cite{Zdunik-2013}, in their comprehensive study for the reconciliation of massive neutron stars in light of the well-known "hyperon puzzle". 

As described in Sec.~\ref{secIII}, the two free parameters are the transition density and the energy density jump. Nevertheless, these properties should be derived by examining the ranges of pressure where each phase is energetically favorable, rather than imposing them by hand. In principle, in the case of AC, one has to evaluate the baryon chemical potential ($\mu_B$) versus pressure relations for each phase, and their intersection indicates the onset of the phase transition. On either side of the transition point, the stable phase is the one that exhibits the higher pressure. In any case, the ansantz described in Eq.~(\ref{1}) ensures that the imposed high-density phase is the stable one for values of chemical potential slightly larger than the transition chemical potential.~However, this does not mean that the high-density phase, as constructed, remains the stable one for higher values of $\mu_B$. In particular, the slope of the $P$-$\mu_B$ curve is the baryon density.~As a consequence the stiffer the EOS, the lower the slope of the curve.~That means that as the chemical potential increases the curve of the imposed high-density phase may pass below the curve of selected low-density phase in the $\mu_B$-$P$ plane (which means that the high-density phase is no longer energetically favored). This is illustrated in Fig.~\ref{instability_pm}, where we depict the baryon chemical potential versus pressure relations for the selected hadronic EOS and maximally stiff EOSs derived through the second line in Eq.~(\ref{1}). The chemical potential is derived as described in Refs.~\cite{Alford-2013,Zdunik-2013}.~As one can observe, the lower the transition density and the energy density jump, the earlier the instability appears (in terms of baryon chemical potential).

Considering the aforementioned issue, we wish to clarify which of the employed parametrizations result into such a behavior. Therefore, we constructed the $\mu_B$-$P$ relations for all of the employed high-density EOSs. The stability was tested for chemical potential values up to 1307.25 MeV (which is the largest value of $\mu_B$ for which the properties of hadronic matter have been calculated, corresponding to $n=3.5n_0$).~In Fig.~\ref{strongPT} we show the parameter space presented in Fig.~\ref{params-m}(a) but we now also include an indication on which parameter sets lead to the aforementioned problem. As it is evident, the derived parameter space is almost entirely covered by this indication. The latter means that, to avoid the instability, the high-density phase should be softer or the hadronic phase should be stiffer (for large values of chemical potential). 

A final remark is that the aim of this section is not to undermine the validity of the results presented in a wide range of studies that employ this model (including the present one). However, it is important to clarify that there is an assumption being made when the constant speed of sound parametrization is used and it should be clearly stated. More precisely, given the challenges associated with precisely determining the nuclear equation of state at high densities, its behavior beyond the phase transition point is not considered.

\section{Conclusion}~\label{secV}

In the present study, we focused on assessing the effects of different speed of sound bounds on the properties of hybrid stars. More precisely, we considered the causality bound, the conformal limit and the relativistic kinetic theory constraint, to describe the high-density phase as maximally stiff. The main goal was to utilize recent  astronomical observations to derive constraints on the properties of first-order phase transitions in compact stars. For the hadronic phase we employed a state-of-the-art microscopic model, for the purpose of increasing the reliability of the derived constraints. Finally, the construction of hybrid EOSs involved the use of two distinct methods (abrupt and smooth phase transitions). 

By analyzing the dependence of the maximum mass on the phase transition density, we verified the results from previous studies that indicate the decreasing trend of $M_{\rm max}$ with increasing $n_{\rm tr}$. However, we found that the inclusion of a softening in the EOS (an energy density jump in the case of an abrupt construction and a crossover in the case of a smooth construction) may alter the aforementioned trend. In addition, we verified the result from previous studies~\cite{Altiparmak-2022,Brandes-2024,Alford-2014b} which indicate that the conformal bound is inconsistent with the existence of massive compact stars (that exceed two $2M_\odot$) when a nonzero energy density jump value (or a transition density above $2n_0$) is considered. 

In the next step, we focused on extracting constraints on the $\Delta\mathcal{E}-n_{\rm tr}$ parameter space by utilizing different astronomical observations. The key idea was that the simultaneous reconciliation of the HESS J1731-347 and PSR J0030+0451 constraints (in $1\sigma$) requires stiffening of the EOS at intermediate density values. Hence, it was crucial to investigate how the consideration of different speed of sound bounds affects the ability of hybrid models to explain the two previously mentioned measurements. In addition, the largest possible energy density jump for a given transition density is driven by the lowest estimation on the PSR J0030+0451 radius. Therefore, the description of the high-density phase as maximally stiff allowed the determination of quite general upper bounds in $\Delta\mathcal{E}$ for each respective speed of sound limit. 

For the speed of sound bounds imposed by special theory of relativity and relativistic kinetic theory, we found that the parameter sets that satisfy the HESS J1731-347 and PSR J0030+0451 constraints are quite similar. Furthermore, the location and range of the derived parameter spaces were found to be sensitive to the selection of the construction method (depending on $\Gamma_m$). For the case of AC, we found that the transition density is constrained below 0.28 fm$^{-3}$, while the energy density jump should be lower than $\sim170$ MeV fm$^{-3}$. The results were similar to the case of SC with $\Gamma_m=1.03$. However, our analysis indicated that increasing the value of the polytropic index reduces the maximum possible transition density and increases the viable energy density jump values. Finally, we showed that the simultaneous explanation of the aforementioned observations is impossible if the sound velocity is bound below $c/\sqrt{3}$.

Lastly, we discussed that the modeling of the high-density phase as maximally stiff may lead to a thermodynamical inconsistency.~More precisely, the hadronic phase may become energetically favored in the region where matter is described using the constant speed of sound parametrization. That is of particular importance as it means that the EOS of the high-density phase should either soften for large chemical potential values (which would affect the resulting maximum mass) or that the hadronic phase is actually stiffer than described (at high densities).

\begin{acknowledgements}
The research work was supported by the Hellenic Foundation for Research and Innovation (HFRI) under the 5th Call for HFRI PhD Fellowships (Fellowship No.~19175).~This work was partly
supported by the Deutsche Forschungsgemeinschaft (DFG, German Research Foundation) under Germany's Excellence Strategy EXC-2094-390783311, ORIGINS and by the National Natural Science Foundation of China (Grants No. 12205030, No. 11935003, and No. 12435006), the National Key Laboratory of Neutron Science and Technology under Grant No. NST202401016, and the Project No. 2024CDJXY022 supported by the Fundamental Research Funds for the Central Universities.
\end{acknowledgements}


\begin{thebibliography}{99}

\bibitem{Bielich-2020} J. Schaffner-Bielich, {\it Compact Star Physics} (Cambridge University Press, Cambridge, UK, 2020).
\bibitem{Lattimer-2001} J. M. Lattimer and M. Prakash, Astrophys. J. {\bf550}, 426 (2021).
\bibitem{Page-2004} D. Page, M. Prakash, James M. Lattimer, A. W. Steiner, 	Astrophys. J. Suppl. {\bf155}, 623 (2004).
\bibitem{Page-2011} D. Page, J. M. Lattimer, M. Prakash, A. W. Steiner, Phys. Rev. Lett. {\bf106}, 081101 (2011).
\bibitem{Lyra-2023} F. Lyra, L. Moreira, R. Negreiros, R. O. Gomes, V. Dexheimer, Phys. Rev. C {\bf 107}, 025806 (2023)
\bibitem{Flores-2014} C. V. Flores, G. Lugones, Class. Quantum Grav. {\bf 31}, 15502 (2014).
\bibitem{Ranea-2018} I. F. Ranea-Sandoval, O. M. Guilera, M. Mariani and M. G. Orsaria, JCAP {\bf12}, 31 (2018). 
\bibitem{Pradhan-2021} B. K. Pradhan, D. Chatterjee, Phys.Rev.C {\bf103}, 035810 (2021).
\bibitem{Pradhan-2024} B. K. Pradhan, D. Chatterjee, D. E. Alvarez-Castillo, Mon. Not. Roy. Astron. Soc. {\bf531}, 4640 (2024).
\bibitem{Moustakidis-2015} Ch.C. Moustakidis, Phys. Rev. C {\bf91}, 035804 (2015).
\bibitem{Riley-2019} T. E. Riley, A. L. Watts, S. Bogdanov, P. S. Ray, R. M. Ludlam,
S. Guillot, Z. Arzoumanian, C. L. Baker, A. V. Bilous, D.
Chakrabarty, K. C. Gendreau, A. K. Harding, W. C. G. Ho, J. M. Lattimer, S. M. Morsink, and T. E. Strohmayer, Astrophys. J. Lett {\bf887}, L21 (2019).
\bibitem{Miller-2019} M. C. Miller, F. K. Lamb, A. J. Dittmann, S. Bogdanov, Z.
Arzoumanian, K. C. Gendreau, S. Guillot, A. K. Harding, W.
C. G. Ho, J. M. Lattimer, R. M. Ludlam, S. Mahmoodifar, S.
M. Morsink, P. S. Ray, T. E. Strohmayer, K. S. Wood, T. Enoto,
R. Foster, T. Okajima, G. Prigozhin {\it et al.}, Astrophys. J. Lett {\bf887}, L24 (2019).
\bibitem{Raaijmakers-2019} G. Raaijmakers, T. E. Riley, A. L. Watts, S. K. Greif, S. M.
Morsink, K. Hebeler, A. Schwenk, T. Hinderer, S. Nissanke, S.
Guillot, Z. Arzoumanian, S. Bogdanov, D. Chakrabarty, K. C.
Gendreau, W. C. G. Ho, J. M. Lattimer, R. M. Ludlam, and M.
T. Wolff, Astrophys. J. Lett {\bf887}, L22 (2019).
\bibitem{Vinciguerra-2024} S. Vinciguerra, T. Salmi, A. L. Watts, D. Choudhury, Th.
E. Riley, P. S. Ray, S. Bogdanov, Y. Kini, S. Guillot, D.
Chakrabarty, W. C. G. Ho, D. Huppenkothen, S. M. Morsink,
Z. Wadiasingh, and M. T. Wolff, Astrophys. J. {\bf 961}, 62 (2024).
\bibitem{Doroshenko-2022} V. Doroshenko, V. Suleimanov, G. Phlhofer, and Andrea
Santangelo, Nat. Astron. {\bf 6}, 1444 (2022).
\bibitem{Akmal-1998} A. Akmal, V. R. Pandharipande, and D. G. Ravenhall
Phys. Rev. C {\bf 58}, 1804 (1998)
\bibitem{Gezerlis-2014} A. Gezerlis, I. Tews, E. Epelbaum, M. Freunek, S. Gandolfi, K. Hebeler, A. Nogga, and A. Schwenk
Phys. Rev. C {\bf 90}, 054323 (2014)
\bibitem{Lynn-2016} J.E. Lynn, I. Tews, J. Carlson, S. Gandolfi, A. Gezerlis, K.E. Schmidt, and A. Schwenk, Phys. Rev. Lett. {\bf 116}, 062501 (2016).
\bibitem{Tews-2018} I. Tews, J. Carlson, S. Gandolfi, and S. Reddy {\it et al.}, Astrophys. J. {\bf 860}, 149 (2018)
\bibitem{Suwa-2018} Y. Suwa, T. Yoshida, M. Shibata, H. Umeda, and K. Takahashi, Mon. Not. Roy. Astron. Soc. {\bf481}, 3305 (2018).
\bibitem{DiClemente-2023} F. Di Clemente, A. Drago, G. Pagliara, Astrophys. J {\bf967}, 159 (2024).
\bibitem{Horvath-2023} J. E. Horvath, L. S. Roch, L. M. de Sá, P. H. R. S. Moraes, L. G. Barão, M. G. B. de Avellar, A. Bernardo and R. R. A. Bachega, Astron. Astrophys. {\bf672}, L11 (2023).
\bibitem{Oikonomou-2023} P.T. Oikonomou and Ch.C. Moustakidis
Phys. Rev. D {\bf 108}, 063010 (2023).
\bibitem{Das-2023} H. C. Das, Luiz L. Lopes, Mon. Not. Roy. Astron. Soc. {\bf525}, 3571 (2023).
\bibitem{Rather-2023} I. A. Rather, G. Panotopoulos, and I. Lopes, Eur. Phys. J. C {\bf83}, 1065 (2023).
\bibitem{Tsaloukidis-2023} L. Tsaloukidis, P.S. Koliogiannis, A. Kanakis-Pegios, and Ch.C. Moustakidis
Phys. Rev. D {\bf 107}, 023012 (2023).
\bibitem{Brodie-2023} L. Brodie and A. Haber,
Phys. Rev. C {\bf 108}, 025806 (2023).
\bibitem{Sagun-2023} V. Sagun, E. Giangrandi, T. Dietrich, O. Ivanytskyi, R. Negreiros,
and C. Providencia, Astrophys. J. {\bf 958}, 49 2023 (2023).
\bibitem{Huang-2023} Kaixuan Huang, Hong Shen, Jinniu Hu, Ying Zhang, Phys. Rev. D {\bf109}, 043036 (2024).
\bibitem{Li-2023} J. J. Li, A. Sedrakian, Phys Lett. {\bf 844B}, 138062 (2023).
\bibitem{Kubis-2023} Sebastian Kubis, Wlodzimierz Wojcik, David Alvarez Castillo, and Noemi Zabari, Phys. Rev. C {\bf108}, 045803 (2023).
\bibitem{Routaray-2023} Pinku Routaray, H. C. Das, Jeet Amrit Pattnaik, Bharat Kumar, arXiv:2307.12748 [math.NA] (2023).
\bibitem{Laskos-2024} P. Laskos-Patkos, P.S. Koliogiannis, Ch.C. Moustakidis, Phys. Rev. D {\bf 109}, 063017 (2024)
\bibitem{Li-2023n} J. J. Li, A. Sedrakian, and M. Alford, Astrophys. J. {\bf967}, 116 (2024).
\bibitem{Mariani-2024} M. Mariani, I. F. Ranea-Sandoval, G. Lugones, M. G. Orsaria, Phys. Rev. D {\bf 110}, 043026 (2024).
\bibitem{Char-2024} P. Char and B. Biswas, 	arXiv:2408.15220 [astro-ph.HE] (2024).
\bibitem{Tewari-2024} S. Tewari, S. Chatterjee, D. Kumar, R. Mallick, 	arXiv:2410.20355 [astro-ph.HE] (2024).
\bibitem{Alford-2023} J. A. J. Alford and J. P. Halpern,  Astrophys. J. {\bf944}, 36 (2023).
\bibitem{Hartle-1978} J. B. Hartle, Phys. Rep. {\bf46}, 201 (1978).
\bibitem{Lattimer-2014} J. M. Lattimer, Astrophysics and Cosmology: Proceedings of the
26th Solvay Conference on Physics, edited by R. Blandford, D.
Gross, and A. Sevrin (World Scientific, Singapore, 2014).
\bibitem{Bedaque-2015} P. Bedaque and A. W. Steiner, Phys. Rev. Lett. {\bf 114}, 031103
(2015).
\bibitem{Olson-2000} T. S. Olson, Phys. Rev. C {\bf 63}, 015802 (2000).
\bibitem{Douchin-2001} F. Douchin and P. Haensel, Astron. Astrophys. {\bf380}, 151-167 (2001).
\bibitem{Hippert-2024} M. Hippert, J. Noronha, P. Romatschke, arXiv:2402.14085 [nucl-th] (2024).
\bibitem{Thompson1970-PRD1-110} R. H. Thompson, Phys. Rev. D {\bf 1}, 110 (1970).
\bibitem{Machleidt1989_ANP19-189} R. Machleidt, Adv. Nucl. Phys. {\bf 19}, 189 (1989).
\bibitem{Brockmann1990_PRC42-1965} R. Brockmann, R. Machleidt, Phys. Rev. C {\bf 42}, 1965 (1990)
\bibitem{Anastasio1980_PRL45-2096} M. R. Anastasio, L. S. Celenza, C. M. Shakin, Phys. Rev. Lett. {\bf 45}, 2096 (1980).
\bibitem{Anastasio1981_PRC23-2273} M. R. Anastasio, L. S. Celenza, C. M. Shakin, Phys. Rev. C {\bf 23}, 2273 (1981).
\bibitem{Anastasio1983_PR100-327} M. R. Anastasio, L. S. Celenza, W. S. Pong, C. M. Shakin, Phys. Rep. {\bf 100}, 327 (1983).
\bibitem{Song1998_PRL81-1584} H. Q. Song, M. Baldo, G. Giansiracusa, U. Lombardo, Phys. Rev. Lett. {\bf 81}, 1584 (1998).
\bibitem{Brockmann1978_PRC18-1510} R. Brockmann, Phys. Rev. C {\bf 18}, 1510 (1978).
\bibitem{WANG-SB2021_PRC103-054319} S.B. Wang, Q. Zhao, P. Ring, J. Meng, Phys. Rev. C {\bf 103}, 054319 (2021).
\bibitem{Wang-2022a} S.B. Wang, H. Tong, Q. Zhao, Ch. Wang, P. Ring, and J, Meng, Phys. Rev. C {\bf 106}, L021305 (2022)
\bibitem{Tong-2022} H. Tong, Ch. Wang, and S.B. Wang, Astrophys. J. {\bf 930}, 137 (2022).
\bibitem{Wang-2022b} S.B. Wang, Ch. Wang, and H. Tong,
Phys. Rev. C {\bf 106}, 045804 (2022).
\bibitem{Qu-2023} X. Qu, H. Tong, H. Tong, Ch. Wang, S.B. Wang, Sci China-Phys. Mech. Astron. {\bf 66}, 242011 (2023).
\bibitem{Qin-2023} P. Qin, Zh. Bai, S.B. Wang, Ch. Wang, and S. Qin,
Phys. Rev. D {\bf107}, 103009 (2023).
\bibitem{Farrell-2024} D. Farrell, F. Weber, Astrophys. J. {\bf969}, 49 (2024). 
\bibitem{Moustakidis-2016} Ch.C. Moustakidis, T. Gaitanos, Ch. Margaritis, G.A. Lalazissis, Phys. Rev. C {\bf 95}, 045801 (2017)
\bibitem{Margaritis-2020} Ch. Margaritis, P.S. Koliogiannis, Ch.C. Moustakidis, Phys. Rev. D {\bf 101}, 043023 (2020).
\bibitem{Kanakis-2020} A. Kanakis-Pegios, P.S. Koliogiannis, Ch.C. Moustakidis, Phys. Rev. C {\bf102}, 055801 (2020).
\bibitem{Reed-2020} B. Reed and C. J. Horowitz, Phys. Rev. C 101, 045803 (2020).
\bibitem{Nath-2002} N. R. Nath, T. E. Strohmayer, and J. H. Swank, Astrophys. J. {\bf564}, 353 (2002).
\bibitem{Roy-2024} S. Roy and 
T. Suyama, Res. in Phys. {\bf61}, 107757 (2024).
\bibitem{Glendenning-1992} N. K. Glendenning
Phys. Rev. D {\bf46}, 1274 (1992).
\bibitem{Costantinou-prd} C. Constantinou, T. Zhao, S. Han, M. Prakash, Phys. Rev. D {\bf107}, 074013 (2023).
\bibitem{Alford-2013} M. G. Alford, S. Han, and M. Prakash, Phys. Rev. D {\bf 88}, 083013 (2013).
\bibitem{Montana-2019} G. Montaña, L. Tolós, M. Hanauske, and L. Rezzolla
Phys. Rev. D {\bf 99}, 103009 (2019).
\bibitem{Antoniadis-2013} J. Antoniadis, P. C. C. Freire, N. Wex, T. M. Tauris, R. S. Lynch,
M. H. van Kerkwijk, M. Kramer, C. Bassa, V. S. Dhillon, T.
Driebe, J. W. T. Hessels, V. M. Kaspi, V. I. Kondratiev, N.
Langer, T. R. Marsh, M. A. McLaughlin, T. T. Pennucci, S. M.
Ransom, I. H. Stairs, J. van Leeuwen, Science {\bf 340}, 448
(2013).
\bibitem{Cromatie-2020} H. T. Cromartie, E. Fonseca, S. M. Ransom, P. B. Demorest,
Z. Arzoumanian, H. Blumer, P. R. Brook, M. E. DeCesar, T.
Dolch, J. A. Ellis, R. D. Ferdman, E. C. Ferrara, N. GarverDaniels, P. A. Gentile, M. L. Jones, M. T. Lam, D. R. Lorimer,
R. S. Lynch, M. A. McLaughlin, C. Ng, Nat. Astron. {\bf 4},
72 (2020). 
\bibitem{Romani-2022} R. G. Romani, D. Kandel, A. V. Filippenko, T. G. Brink, and W. Zheng, Astrophys. J. Lett. {\bf 934}, L17 (2022).
\bibitem{Altiparmak-2022} S. Altiparmak, C. Ecker, L. Rezzolla, Astrophys.J.Lett. {\bf939}, L34 (2022).
\bibitem{Ecker-2022} C. Ecker, L. Rezzolla, Astrophys.J.Lett. {\bf939}, L35 (2022).
\bibitem{Brandes-2024}
L. Brandes, W. Weise, Symmetry {\bf16(1)}, 111 (2024).
\bibitem{Sun-2023} H. Sun and D. Wen
Phys. Rev. C {\bf 108}, 025801 (2023).
\bibitem{Alford-2014b} M. G. Alford, G. F. Burgio, S. Han, G. Taranto, and D. Zappalà, Phys. Rev. D {\bf 92}, 083002 (2015)
\bibitem{Christian-2019} J.E. Christian, A. Zacchi, and  J. Schaffner-Bielich, Phys. Rev. D \textbf{99}, 023009 (2019).
\bibitem{Christian-2021} J.E. Christian and  J. Schaffner-Bielich, Phys. Rev. D \textbf{103}, 063042 (2021)
\bibitem{Christian-2022} J.E. Christian and  J. Schaffner-Bielich, Astrophys. J. {\bf 935}, 122 (2022).
\bibitem{Han-2019a} S. Han and A.W. Steiner, Phys. Rev. D \textbf{99}, 083014 (2019).
\bibitem{Li-2021}J. J. Li, A. Sedrakian, and M. Alford, Phys. Rev. D {\bf 104}, L121302 (2021).
\bibitem{Sharifi-2021}Z. Sharifi, M. Bigdeli, and D. Alvarez-Castillo, Phys. Rev. D {\bf 103}, 103011 (2021).
\bibitem{Paschalidis-2018} V. Paschalidis, K. Yagi, D. Alvarez-Castillo, D.B. Blaschke, and A. Sedrakian, Phys. Rev. D {\bf 97}, 084038 (2018).
\bibitem{Niseri-2024} M. Naseri, G. Bozzola, and V. Paschalidis,
Phys. Rev. D {\bf 110}, 044037 (2024)
\bibitem{Alford-2017} M. Alford and A. Sedrakian, Phys. Rev. Lett. {\bf 119}, 161104 (2017).
\bibitem{Deloudis-2021}	T. Deloudis, P. Koliogiannis, and Ch. Moustakidis, EPJ Web Conf. {\bf 252}, 06001 (2021).
\bibitem{Han-2020}S. Han and M. Prakash, Astrophys. J. {\bf 899}, 164 (2020).
\bibitem{Christian-2024} J.-E. Christian, J. Schaffner-Bielich, S. Rosswog,
Phys. Rev. D {\bf 109}, 063035 (2024).
\bibitem{Li-2023h} J. J. Li, A. Sedrakian, and M. Alford, Phys. Rev. D {\bf 107}, 023018 (2023).
\bibitem{Zdunik-2013} J.L. Zdunik, P. Haensel, Astron. Astrophys. {\bf 551}, A61 (2013).



\end{thebibliography}
\end{document}